\documentclass[11pt]{article}
\usepackage{amssymb}
\usepackage{colortbl}
\usepackage{amsfonts,amsmath, longtable}

        \def\theequation{\thesection.\arabic{equation}}

\newcommand{\tr}{{\rm tr}}
\newcommand{\ti}[1]{\tilde{#1}}

\newcommand{\mH}{{\mathcal H}}

\newcommand{\vf}{\varphi}
\newcommand{\al}{\alpha}
\newcommand{\be}{\beta}

\newcommand{\om}{\omega}
\newcommand{\vth}{\vartheta}

\newcommand{\bfe}{{\bf{e}}}

\newcommand{\Mat}{ {\rm Mat}(N,\mathbb C) }

\newcommand{\mC}{\mathbb C}
\newcommand{\mZ}{\mathbb Z}

\def\beq{\begin{equation}}
\def\eq{\end{equation}}
\def\p{\partial}

\newcommand{\mat}[4]{\left(\begin{array}{cc}{#1}&{#2}\\ \ \\{#3}&{#4}
\end{array}\right)}

\def\res{\mathop{\hbox{Res}}\limits}

\begin{document}

\setcounter{page}{1}

\begin{center}

\

\vspace{0mm}

{\Large{\bf
Field generalization of elliptic Calogero-Moser system
}}

\vspace{3mm}

{\Large{\bf
in the form of higher rank Landau-Lifshitz model
}}

 \vspace{15mm}

 {\Large {K. Atalikov}}$\,^{\bullet}$$\,^{\diamond}$
\qquad\quad\quad
 {\Large {A. Zotov}}$\,^{\diamond}$

  \vspace{10mm}

$\bullet$ -- {\em NRC ''Kurchatov Institute'',
Kurchatova sq. 1, 123182, Moscow, Russia}

$\diamond$ -- {\em Steklov Mathematical Institute of Russian
Academy of Sciences,\\ Gubkina str. 8, 119991, Moscow, Russia}



   \vspace{3mm}

 {\small\rm {e-mails: kantemir.atalikov@yandex.ru, zotov@mi-ras.ru}}

\end{center}

\vspace{0mm}

\begin{abstract}
We prove gauge equivalence between integrable field generalization of the elliptic Calogero-Moser model
and the higher rank XYZ Landau-Lifshitz model of vector type on 1+1 dimensional space-time.
Explicit formulae for the change of variables are derived, thus providing the Poisson map
between these models.
\end{abstract}


%


\bigskip


\section{Introduction}\label{sec1}
\setcounter{equation}{0}

We consider two types of integrable field theories on 1+1 dimensional space-time.
The first one is the Landau-Lifshitz model \cite{LL}, describing behaviour
of the magnetization vector $\vec S(t,x)=(S_1,S_2,S_3)$ in the one-dimensional
model of ferromagnet:
 \beq\label{aa01}
 \begin{array}{c}
   \displaystyle{
 \partial_{t} {\vec S}={\vec S}\times{\vec S}_{xx}+{\vec S}\times J({\vec S})\,,
 \quad {\vec S}_{xx}=\p_x^2{\vec S}\,,
 }
 \end{array}
  \eq
  where $J({\vec S})=(J_1S_1,J_2S_2,J_3S_3)$ with some constants $J_1,J_2,J_3$ describing
  the anisotropy.
 Here $t\in{\mathbb R}$ is the time variable, and $x$ is the space variable.
We assume $x$ be a coordinate on a unit circle, and all the fields in this paper are
periodic $\psi(t,x)=\psi(t,x+2\pi)$. We also imply that all the fields are $\mC$-valued.
Integrability of
this model was proved in \cite{Skl} through the classical inverse scattering method \cite{FT,ZaSh}.
In particular, it was shown that the equation (\ref{aa01}) is represented in the
for of the Zakharov-Shabat equation (or, the Lax equation, or the zero curvature condition):
 \beq\label{aa02}
 \begin{array}{c}
   \displaystyle{
    \partial_{t} U(z)-k \partial_{x} V(z)+[U(z), V(z)]=0\,,\quad \forall z\,,
     }
 \end{array}
   \eq
where $U(z),V(z)\in{\rm Mat}_2$ are $2\times 2$ matrices,  $z$ is a complex valued spectral parameter
and $k\in\mathbb R$ is a parameter.

The second model is the field generalization of 2-body elliptic Calogero-Moser system \cite{Krich2,LOZ}.
 At the level of classical finite-dimensional mechanics the 2-body system is described by the
 Hamiltonian
 \beq\label{aa03}
 \begin{array}{c}
   \displaystyle{
 H^{\hbox{\tiny{CM}}}=\frac{p^2}{2} - \frac{c^2}{8}\wp(q)
 }
 \end{array}
  \eq
and the canonical Poisson bracket $\{p,q\}=1$ between the momentum $p$ and the position
of particle\footnote{In fact, there is a pair of particles. The Hamiltonian (\ref{aa03}) is written
in the center of mass frame, so that $q=q_1-q_2$. For this reason the normalization of the Hamiltonian
slightly differs in 2-body case compared to $N$-body case.} $q$. In (\ref{aa03}) $c\in\mC$ is a coupling constant
and $\wp(q)$ is the elliptic Weierstrass $\wp$-function. It is an elliptic version for the inverse square function.
All necessary definition of elliptic functions are given in the Appendix.
In the field theory the Hamiltonian takes the form \cite{Krich2,LOZ}
 \beq\label{aa04}
 \begin{array}{c}
   \displaystyle{
 \mH^{\hbox{\tiny{2dCM}}}= \frac{1}{2}\oint dx \left( {{p}^{2}}\Big( 1-\frac{{{k}^{2}}q_{x}^{2}}{{{c}^{2}}} \Big)+\frac{(3{{k}^{2}}q_{x}^{2}-{{c}^{2}})}{4}\wp(q)-\frac{{{k}^{4}}{{q}^{2}_{xx}}}{4(c^2-k^2q_x^2)}\right)\,,
 }
 \end{array}
  \eq
  where $c$ is the coupling constant and $\zeta(u)$ the elliptic Weierstrass $\zeta$-function, see Appendix.
  In this model we deal with the canonical fields $p(x)$ and $q(x)$ on a unit circle: $\{p(x),q(y)\}=\delta(x-y)$.

  It was explained in \cite{LOZ} and then computed in \cite{AtZ2} that the Landau-Lifshitz model is gauge equivalent
  to the field Calogero-Moser system. This means that
  there exists a matrix $G(z)\in{\rm Mat}_2$, which relates $U$-matrices of both models through the gauge
  transformation:
 \beq\label{aa05}
 \begin{array}{c}
   \displaystyle{
U^{\hbox{\tiny{LL}}}(z)=G(z) U^{\hbox{\tiny{2dCM}}}(z) G^{-1}(z)+k G_x(z) G^{-1}(z)\,.
 }
 \end{array}
  \eq
  The matrix $G(z)$ depends on dynamical variables. It is a continuous version of the IRF-Vertex transformation
  introduced by R.J. Baxter \cite{Baxter2} for the quantum statistical models. In this treatment the Landau-Lifshitz model is of the vertex type, while
  the Calogero-Moser system is on the IRF (interacting round a face) side.

  \paragraph{Purpose of the paper.} In this paper we generalize the above results to the higher rank models. The ${\rm gl}_N$
  Calogero-Moser model \cite{Calogero2,Kr} in classical mechanics is described by $N$-body Hamiltonian
  \beq\label{aa201}
  \begin{array}{c}
  \displaystyle{
H^{\hbox{\tiny{CM}}}=\sum\limits_{i=1}^N
\frac{p_i^2}{2}-{c}^2\sum\limits_{i>j}^N\wp(q_i-q_j)\,,
 }
 \end{array}
 \eq
  Its field generalization was proposed in \cite{Krich22} using reduction from
  matrix KP equations, and the integrability was proved in \cite{Z24}
  through the classical $r$-matrix structure. We describe this model in detail
  in Section \ref{sec3}. The higher rank generalization if the Landau-Lifshitz model
  was derived in \cite{AtZ4} through the associative Yang-Baxter equation.
  It is also described in Section \ref{sec3}. Two models
  are related by the gauge transformation as given in (\ref{aa05}). The matrix of the corresponding
  gauge transformation is a continuous version of the IRF-Vertex transformation
  for Belavin's $R$-matrix found in \cite{Jimbo}. Moreover, in \cite{Z24}
  the continuous version of the IRF-Vertex transformation was performed at the
  level of classical $r$-matrix structures for both field theories.
  Main purpose of this paper is to finish description of the gauge equivalence by
  evaluating explicit change of variables relating both integrable field theories.
  Similar results for the rational and trigonometric models were obtained in our
  previous papers \cite{AtZ2,AtZ3}.

  The paper is organized as follows. In the next Section we recall main results for ${\rm gl}_2$
  models from \cite{AtZ1}. The pair of models in the higher rank case are described in Section \ref{sec3}.
  In Section \ref{sec4} we calculate explicit change of variables and argue that is provides
  the Poisson map between the models. Definitions and properties of elliptic functions
  are given in the Appendix.

\section{An overview of 2-body case}\label{sec2}
\setcounter{equation}{0}

Here we briefly recall the result of \cite{AtZ1}.
Namely , we describe the field analogue of 2-body Calogero-Moser model
and represent it in the form of the Landau-Lifshitz magnet.

\paragraph{Classical mechanics.} The 2-body Calogero-Moser model is described by the Hamiltonian
(\ref{aa03}). Equations of motion take the form
 \beq\label{aa101}
 \begin{array}{c}
   \displaystyle{
 \dot{p}=\frac{c ^2}{8} \wp'(q), ~~~ \dot{q}=p\,.
 }
 \end{array}
  \eq
  They are represented in the Lax form
 \beq\label{aa102}
 \begin{array}{c}
   \displaystyle{
 \dot{L}^{\hbox{\tiny{CM}}}(z)\equiv\{H^{\hbox{\tiny{CM}}}, L^{\hbox{\tiny{CM}}}(z)\}
 =[L^{\hbox{\tiny{CM}}}(z), {M}^{\hbox{\tiny{CM}}}(z)]
 }
 \end{array}
  \eq
with the Lax pair
 \beq\label{aa103}
 \begin{array}{c}
   \displaystyle{
 L^{\hbox{\tiny{CM}}}(z)=\left(\begin{array}{cc}
p & \displaystyle{\frac{c}{2}\, \phi(-z, q) }\\
\displaystyle{ \frac{c}{2}\, \phi(-z, -q) } & -p
\end{array}\right),
 }
 \quad
  \displaystyle{
 M^{\hbox{\tiny{CM}}}(z)=\frac{1}{4}\left(
\begin{array}{cc}
 0 & c  f(-z,q) \\
  c  f(-z,-q) & 0 \\
\end{array}
\right),
 }
 \end{array}
  \eq
where the functions $\phi$ and $f$ are given in the Appendix in (\ref{a01}) and (\ref{a04})
respectively.

 \paragraph{1+1 field theory.} In this case the momentum and coordinate become fields
on a unit circle ${\mathbb S}^1$, and the canonical Poisson bracket  turns into
 \beq\label{aa104}
 \begin{array}{c}
   \displaystyle{
 \left\{p(x), q(y)\right\}= \delta(x-y)\,.
 }
 \end{array}
  \eq
 Equations of motion takes the form:
 \beq\label{aa105}
 \begin{array}{c}
   \displaystyle{
 q_t={p}\left( 1-\frac{{{k}^{2}}q_{x}^{2}}{{{c}^{2}}} \right)\,,
 }
 \\ \ \\
    \displaystyle{
  p_t=-\frac{ k^2}{ c^2} \partial_{x}\left(p^{2} q_{x}\right)-\frac{(3{{k}^{2}}q_{x}^{2}-{{c}^{2}})}{8}\wp'(q)+\frac{3 k^2}{4} \partial_{x}\left( q_{x} \wp(q) \right)+\frac{k^4}{4} \partial_{x}\left(\frac{q_{x x x} {\tilde\nu}-{\tilde\nu}_{x} q_{x x}}{{\tilde\nu}^{3}}\right)\,,
  }
  \end{array}
   \eq
where
  %
 \beq\label{aa106}
 \begin{array}{c}
   \displaystyle{
\tilde\nu=\sqrt{c^2-k^2q_x^2}\,,\quad c={\rm const}\,.
  }
  \end{array}
   \eq
%

   %
 \beq\label{aa107}
 \begin{array}{c}
   \displaystyle{
 U^{\hbox{\tiny{2dCM}}}(z)=\frac12\mat{2p-kq_xE_1(z)}{\tilde\nu\phi(-z,q)}{\tilde\nu\phi(-z,-q)}{-2p+kq_xE_1(z)}
 }
 \end{array}
  \eq

\paragraph{Landau-Lifshitz magnet.}  By inserting three components of $\vec S(t,x)$
into the traceless $2\times 2$ matrix $S=\sum_{\al=1}^3\sigma_\al S_\al$ in the Pauli matrices
$\sigma_\al$ basis, the Landau-Lifshitz equation takes the form:
 \beq\label{aa110}
 \begin{array}{c}
   \displaystyle{
 \partial_{t} S=[J(S), S]-\alpha_0\left[S, S_{x x}\right]\,,\quad S_{xx}=\p_x^2S\,,
 }
 \end{array}
  \eq
 where $J(S)$ is the linear map describing the anisotropy in the magnet
 \beq\label{aa1101}
 \begin{array}{c}
   \displaystyle{
J(S)=\sum\limits_{\al=1}^3 S_\al J_\al\sigma_\al\,,\quad J_\al=-\frac{1}{2}\,\wp(\om_\al)
 }
 \end{array}
  \eq
and
 \beq\label{aa111}
 \begin{array}{c}
   \displaystyle{
\alpha_0=k^2/(8\lambda^2)
 }
 \end{array}
  \eq
is a constant parameter. Here $\lambda\in\mC$ (also constant) is an eigenvalue of the matrix $S$ -- the norm of the vector $(S_1,S_2,S_3)$, and $k$ is the constant coefficient behind $\p_x$.

The Landau-Lifshitz equation (\ref{aa110}) has the Hamiltonian formulation with the Poisson brackets
 \beq\label{aa113}
 \begin{array}{c}
   \displaystyle{
 \left\{S_{\alpha}(x), S_{\beta}(y)\right\}=-\sqrt{-1} \varepsilon_{\alpha \beta \gamma} S_{\gamma}(x)\delta (x-y)\,,
 }
 \end{array}
  \eq
and the Hamiltonian
 \beq\label{aa114}
 \begin{array}{c}
   \displaystyle{
  \mH^{\hbox{\tiny{LL}}}=\frac{1}{2}\oint dx \Big( \tr(SJ(S))-\al_0\tr(S_x^2) \Big)\,.
 }
 \end{array}
  \eq
The $U$-matrix (it is $2\times 2$ matrix) entering the Zakharov-Shabat equation has the form \cite{Skl,FT,STS}:
 \beq\label{aa116}
 \begin{array}{c}
   \displaystyle{
    U^{\hbox{\tiny{LL}}}(z)=\sum\limits_{k=1}^3 S_k\vf_k(z)\sigma_k\,,
     }
 \end{array}
   \eq
where  $\vf_1(z)=e^{\pi\imath z}\phi(z,\frac{\tau}{2})$,
  $\vf_1(z)=e^{\pi\imath z}\phi(z,\frac{1+\tau}{2})$, $\vf_3(z)=\phi(z,\frac{1}{2})$.

\paragraph{Gauge equivalence.} The gauge transformation (\ref{aa05}) with the matrix
 \beq\label{aa117}
 \begin{array}{c}
   \displaystyle{
 G(z)=\frac{1}{\rho}\left(
\begin{array}{cc}
 \theta_{00} (z+q \,|\, 2 \tau ) {\tilde\nu}  & -\theta_{00} (q-z \,|\, 2\tau ) \left(c+k q_{x}\right) \\ \ \\
 -\theta_{10} (z+q \,|\, 2 \tau ) {\tilde\nu}  & \theta_{10} (q-z \,|\, 2 \tau ) \left(c+k q_{x}\right)
\end{array}
\right)\,,
 }
 \end{array}
  \eq
  where $\ti\nu$ is from (\ref{aa106}) and
  \beq\label{aaa118}
  \begin{array}{c}
    \displaystyle{
  \rho=\sqrt{\ti\nu(c+k q_{x})\vartheta(z) \vartheta(q)}\,,
  }
  \end{array}
   \eq
leads to the following change of variables:
 \beq\label{aa120}
 \left\{\begin{array}{l}
   \displaystyle{
   S_{1}(p,q,c) =\Big(p-\frac{c}{2}\,\frac{k^2q_{xx}}{c^2-k^2q_x^2}\Big) \frac{\theta_{01}(0)}{\vartheta^{\prime}(0)} \frac{\theta_{01}(q)}{\vartheta(q)}+\frac{c}{2}\, \frac{\theta_{01}^{2}(0)}{\theta_{00}(0) \theta_{10}(0)} \frac{\theta_{00}(q) \theta_{10}(q)}{\vartheta^{2}(q)}\,,
   }
    \\ \ \\
    \displaystyle{
   S_{2}(p,q,c) =\Big(p-\frac{c}{2}\,\frac{k^2q_{xx}}{c^2-k^2q_x^2}\Big) \frac{\sqrt{-1}\theta_{00}(0)}{ \vartheta^{\prime}(0)} \frac{\theta_{00}(q)}{\vartheta(q)}
   +\frac{c}{2}\, \frac{\sqrt{-1}\theta_{00}^{2}(0)}{ \theta_{10}(0) \theta_{01}(0)}
    \frac{\theta_{10}(q) \theta_{01}(q)}{\vartheta^{2}(q)}\,,
   }
    \\ \ \\
   \displaystyle{
    S_{3}(p,q,c) =\Big(p-\frac{c}{2}\,\frac{k^2q_{xx}}{c^2-k^2q_x^2}\Big) \frac{\theta_{10}(0)}{\vartheta^{\prime}(0)} \frac{\theta_{10}(q)}{\vartheta(q)}+\frac{c}{2}\, \frac{\theta_{10}^{2}(0)}{\theta_{00}(0) \theta_{01}(0)} \frac{\theta_{00}(q) \theta_{01}(q)}{\vartheta^{2}(q)}\,.
 }
 \end{array}
 \right.
  \eq
  In the above formulae the notations (\ref{a121}) are used.

\section{$N$-body case: description of models}\label{sec3}
\setcounter{equation}{0}

\subsection{Field analogue of $N$-body elliptic Calogero-Moser model.}

\paragraph{Hamiltonian and equations of motion.}
The $N$-body Calogero-Moser model in classical mechanics is described by the Hamiltonian (\ref{aa201}).
 Its field generalization was derived in \cite{Krich22}. The Hamiltonian takes the form
 \beq\label{aa202}
 \begin{array}{c}
   \displaystyle{
\mH^{\hbox{\tiny{2dCM}}}= \oint dx\, \mH^{\hbox{\tiny{2dCM}}}(x)
 }
 \end{array}
  \eq
  with the density\footnote{In the limit to the finite-dimensional mechanics all the fields become independent of $x$.
  This corresponds to the limit $k\rightarrow 0$. In this limit $\mH^{\hbox{\tiny{2dCM}}}(x)\rightarrow 2cH^{\hbox{\tiny{CM}}}$.}
 \beq\label{aa203}
 \begin{array}{c}
   \displaystyle{
\mH^{\hbox{\tiny{2dCM}}}(x)= \sum_{i=1}^{N} p_{i}^{2}
\left(c-k q_{i x}\right)-\frac{1}{N c}\Big(\sum_{i=1}^{N} p_{i}\left(c-k q_{i x}\right)\Big)^{2}-
 }
 \\
 \displaystyle{
 -\sum_{i=1}^{N} \frac{k^{4} q_{i x x}^{2}}{4\left(c-k q_{i x}\right)}+\frac{k^{3}}{2} \sum_{i \neq j}^N
 \Big(q_{i x} q_{j x x}-q_{j x} q_{i x x}\Big) \zeta\left(q_{i}-q_{j}\right) -
 }
  \\
 \displaystyle{
-\frac{1}{2} \sum_{i \neq j}^N\Big(\left(c-k q_{i x}\right)^{2}
\left(c-k q_{j x}\right)+\left(c-k q_{i x}\right)
\left(c-k q_{j x}\right)^{2}-c k^{2} \left(q_{i x}-q_{j x}\right)^{2}\Big)  \wp\left(q_{i}-q_{j}\right)\,.
 }
 \end{array}
  \eq
Together with the canonical Poisson structure
 \beq\label{aa2031}
 \begin{array}{c}
   \displaystyle{
\{q_i(x),p_j(y)\}=\delta(x-y)\,,\quad
\{q_i(x),q_j(y)\}=\{p_i(x),p_j(y)\}=0
 }
 \end{array}
  \eq
it provides equations of motion
 \beq\label{aa204}
 \begin{array}{c}
   \displaystyle{
\dot{q}_{i}= 2 p_{i}\left(c-k q_{i x}\right)-\frac{2}{N c} \sum_{l=1}^{N} p_{l}\left(c-k q_{l x}\right)\left(c-k q_{i x}\right)
 }
 \end{array}
  \eq
and
 \beq\label{aa205}
 \begin{array}{c}
   \displaystyle{
\dot{p}_{i}=-2 k p_{i} p_{i x}+k\p_x\Big(
\frac{k^{3} q_{i x x x}}{2\left(c-k q_{i x}\right)}+\frac{k^{4} q_{i x x}^{2}}{4\left(c-k q_{i x}\right)^{2}}+
\frac{2 }{N c}\sum_{l=1}^{N} p_{i} p_{l}\left(c-k q_{l x}\right)
\Big)
+
 }
 \\
  \displaystyle{
+2 \sum_{j:j \neq i}^N\Big(k^{3} q_{j x x x} \zeta \left(q_{i}-q_{j}\right) - 3 k^{2} \left(c-k q_{j x}\right) q_{j x x} \wp \left(q_{i}-q_{j}\right) + \left(c-k q_{j x}\right)^{3}\wp' \left(q_{i}-q_{j}\right) \Big)\,.
 }
 \end{array}
  \eq
\paragraph{$U-V$ pair.} The equations of motion
(\ref{aa205}) are
 represented in the Zakharov-Shabat form (\ref{aa02}) with $N\times N$ matrices
$U^{\hbox{\tiny{2dCM}}}(z)$ and $V^{\hbox{\tiny{2dCM}}}(z)$. The entries of the matrices are as follows:
 \beq\label{aa206}
 \begin{array}{c}
   \displaystyle{
U^{\hbox{\tiny{2dCM}}}_{i j}(z)=\delta_{i j} \Big(-p_i + \frac{1}{N} \sum_{k=1}^{N} p_{k}
-  \alpha_{i}^{2} E_{1}(z)-\frac{k \alpha _{i x}}{\alpha _i}\Big)-\left(1-\delta_{i j}\right)  \alpha_{j}^2  \phi\left(q_{j}-q_{i}, z\right)
     }
 \end{array}
   \eq
and
 \beq\label{aa207}
 \begin{array}{c}
   \displaystyle{
V^{\hbox{\tiny{2dCM}}}_{i j}(z)=\delta_{i j}\Big(-q_{i t} E_{1} \left(  z \right)- N c \alpha_{i}^{2} \wp\left(  z \right)+\widetilde{m}_{i}^{0} -\frac{\alpha_{i t}}{\alpha_{i}}-\frac{1}{N}\sum_{i=1}^N \tilde{m}_{i}^{0}\Big)-
     }
     \\ \ \\
   \displaystyle{
-\left(1-\delta_{i j}\right) \alpha_{j}^{2} \Big( N c E_{1}( z)\phi\left(-q_{i}+q_{j},  z\right) -N c \phi^{\prime}\left(-q_{i}+q_{j}, N z\right)-\widetilde{m}_{i j} \phi\left(-q_{i}+q_{j},  z\right)\Big)\,.
   }
 \end{array}
   \eq
  Here we use notations
 \beq\label{aa208}
 \begin{array}{c}
   \displaystyle{
\alpha_{i}^{2}=k q_{i x}-c\,,
     }
 \end{array}
   \eq
 \beq\label{aa209}
 \begin{array}{c}
   \displaystyle{
\tilde{m}_{i}^{0}=p_{i}^{2}+\frac{k^{2} \alpha_{i x x}}{\alpha_{i}}+2 \kappa p_{i}-
\sum_{k:k \neq i}^N\Big(\left(2 \alpha_{k}^{4}+\alpha_{i}^{2}\alpha_{k}^{2}\right)
\wp\left(q_{i}-q_{k}\right)+4 k \alpha_{k} \alpha_{k x} \zeta\left(q_{i}-q_{k}\right)\Big)\,,
     }
 \end{array}
   \eq
 \beq\label{aa210}
 \begin{array}{c}
   \displaystyle{
\widetilde{m}_{i j}=p_{i}+p_{j}+2 \kappa+\frac{k \alpha_{i x}}{\alpha_{i}}-\frac{k \alpha_{j x}}{\alpha_{j}}-
\sum_{k:k \neq i, j}^N \alpha_{k}^{2} \eta\left(q_{i}, q_{k}, q_{j}\right)\,,
     }
 \end{array}
   \eq
 \beq\label{aa211}
 \begin{array}{c}
   \displaystyle{
\eta(\lambda, \nu, \mu)=\zeta(\lambda-\nu)+\zeta(\nu-\mu)-\zeta(\lambda-\mu)
     }
 \end{array}
   \eq
  and
 \beq\label{aa212}
 \begin{array}{c}
   \displaystyle{
\kappa=-\frac{1}{N c} \sum_{k=1}^{N} p_{k}\left(c-k q_{k x}\right)\,.
     }
 \end{array}
   \eq

\subsection{Higher rank Landau-Lifshitz model}
\paragraph{Notations.}
Introduce
the special matrix basis in $\Mat$:
  \beq\label{aa231}
  \begin{array}{l}
  \displaystyle{
T_a=T_{a_1 a_2}=\exp\left(\frac{\pi\imath}{{ N}}\,a_1
 a_2\right)Q^{a_1}\Lambda^{a_2}\,,
 \qquad
 a=(a_1,a_2)\in\mZ_{ N}\times\mZ_{ N}\,,\quad \mZ_N=\mZ/N\mZ\,,
 }
 \\ \ \\
  \displaystyle{
(Q)_{kl}=\delta_{kl}\exp(\frac{2\pi
 \imath}{{ N}}k)\,,
 \qquad
 (\Lambda)_{kl}=\delta_{k-l+1=0\,{\hbox{\tiny{mod}}}\,
 { N}}\,,\quad k,l=1,..,N\,.
 }
 \end{array}
 \eq
The basis matrices are numerated by a pair of numbers $(a_1,a_2)$, $a_1,a_2=0,...,N-1$ defined modulo $N$.
In particular, $T_{(0,0)}=1_N$ -- identity $N\times N$ matrix. Then
  \beq\label{aa232}
 \begin{array}{c}
  \displaystyle{
T_\al T_\be=\kappa_{\al,\be} T_{\al+\be}\,,\ \ \
\kappa_{\al,\be}=\exp\left(\frac{\pi \imath}{N}(\be_1
\al_2-\be_2\al_1)\right)\,,
 }
 \end{array}
 \eq
The basis has the property $\tr(T_\al T_\be)=N\delta_{\al+\be,(0,0)}$.
 See details in the Appendix of the paper \cite{ZZ}.
Below we use the following set of functions:
 \beq\label{aa233}
 \begin{array}{c}
  \displaystyle{
 \vf_a(z,\om_a+\hbar)=\exp (2\pi\imath\,\frac{a_2z}{N})\,\phi(z,\om_a+\hbar)=
 \exp (2\pi\imath z\p_\tau\om_a)\,\phi(z,\om_a+\hbar)\,,
 \quad a\in\,\mZ_{ N}\times\mZ_{ N}

 }
 \end{array}
 \eq
  where
 \beq\label{aa234}
 \begin{array}{c}
  \displaystyle{
 \om_a=\frac{a_1+a_2\tau}{N}\,,
 \quad a\in\,\mZ_{ N}\times\mZ_{ N}\,.
 }
 \end{array}
 \eq
 The classical elliptic Belavin-Drinfeld $r$-matrix \cite{BD} takes the following simple form:
 \beq\label{aa235}
 \begin{array}{c}
  \displaystyle{
 r_{12}(z)=E_1(z)1_N\otimes 1_N+\sum\limits_{a\neq(0,0)} T_a\otimes T_{-a}
 \exp (2\pi\imath\,\frac{a_2z}{N})\,\phi(z,\frac{a_1+a_2\tau}{N})\,,
 }
 \end{array}
 \eq
where the sum over $a$ goes over all $a\in\mZ_N\times\mZ_N$, $a\neq(0,0)$. We use this notation
in what follows.

Notice that the described above matrix basis reproduces (up to signs) the Pauli matrices
in the $N=2$ case.

The transition between the standard matrix basis and the basis $T_a$ is performed as follows.
Let $B$ be an arbitrary $N\times N$ matrix with entries $B_{ij}$ in the standard basis and
with $B_{(a_1,a_2)}$ components in the basis $T_a$. Then with the short notation
$\bfe(x)=\exp(2\pi\imath x)$ we have
  \beq\label{aa2361}
  \begin{array}{c}
  \displaystyle{
B_a=B_{(a_1,a_2)}=\frac{1}{N}\,\tr(B T_{-a})=\frac{1}{N}\,\bfe(-\frac{a_1a_2}{2N})\sum\limits_{k=1}^N B_{k,k+a_2}\,
\bfe(-\frac{a_1 k}{N})
 }
 \end{array}
 \eq
and
 \beq\label{aa2362}
 \begin{array}{c}
 B_{ij}=\left\{
 \begin{array}{l}
   \displaystyle{
   \sum\limits_{a_1=0}^{N-1}B_{(a_1,j-i)}\bfe\Big(\frac{a_1(i+j)}{2N}\Big)\,,\quad j\geq i\,,
   }
 \\
   \displaystyle{
    \sum\limits_{a_1=0}^{N-1}B_{(a_1,j-i+N)}\bfe\Big(\frac{a_1(i+j-N)}{2N}\Big)\,,\quad j<i\,.
   }
 \end{array}
 \right.
 \end{array}
 \eq

\paragraph{Equations of motion and Hamiltonian formulation.} In ${\rm gl}_N$ case
the Landau-Lifshitz equation (\ref{aa110}) is generalized as \cite{AtZ4}:
 \beq\label{aa237}
 \begin{array}{c}
  \displaystyle{
\partial_{t} S = 2 c [S, J(S)] + \frac{k^{2}}{c} [S, S_{xx}]-2 k [S, E(S_{x})]\,,
 }
 \end{array}
 \eq
 where $S=S(t,x)\in\Mat$ is a matrix of dynamical variables (fields). Here we assume this matrix
 is not an arbitrary but the one, which has a special set of eigenvalues $0,..,0,c$:
 \beq\label{aa238}
 \begin{array}{c}
  \displaystyle{
{\rm Spec}(S)=(0,...,0, c), ~~~\tr(S)= c\,.
 }
 \end{array}
 \eq
 The latter means that $S$ is a rank one matrix. It satisfies the property
 \beq\label{aa2381}
 \begin{array}{c}
  \displaystyle{
S^2=  c S\,.
 }
 \end{array}
 \eq
 The linear maps $J(S)$ and $E(S)$ entering equation (\ref{aa237}) are of the following form:
 \beq\label{aa2382}
 \begin{array}{c}
  \displaystyle{
J(S)=\frac{N \vth'''(0)}{3\vth'(0)}1_N S_{(0,0)}-N \sum\limits_{a\neq(0,0)} T_a S_a
 E_2\Big(\frac{a_1+a_2\tau}{N}\Big)=
 }
 \\
  \displaystyle{
 =\frac{N \vth'''(0)}{3\vth'(0)}\,S-N \sum\limits_{a\neq(0,0)} T_a S_a
 \wp\Big(\frac{a_1+a_2\tau}{N}\Big)
 }
 \end{array}
 \eq
 and
 \beq\label{aa2383}
 \begin{array}{c}
  \displaystyle{
E(S)=N \!\sum\limits_{a\neq(0,0)} \!T_a S_a\Big(2\pi\imath\,
  \frac{a_2}{N}+E_1\Big(\frac{a_1\!+\!a_2\tau}{N}\Big) \Big)\,.
 }
 \end{array}
 \eq
 Notice that in the $N=2$ case $E(S)=0$ for any matrix $S$ and the original Landau-Lifshitz
 equation (\ref{aa110}) is reproduced.

 The equation (\ref{aa237}) has the Hamiltonian description, that is
 \beq\label{aa239}
 \begin{array}{c}
  \displaystyle{
\partial_{t} S(t,x) = \{S(t,x), \mH^{\hbox{\tiny{LL}}} \}
 }
 \end{array}
 \eq
 with the following Hamiltonian
 \beq\label{aa240}
 \begin{array}{c}
  \displaystyle{
\mH^{\hbox{\tiny{LL}}}=\oint d x \mH^{\hbox{\tiny{LL}}}(x)\,,
 }
 \end{array}
 \eq
 \beq\label{aa241}
 \begin{array}{c}
  \displaystyle{
  \mH^{\hbox{\tiny{LL}}}(x)=
N c \tr(S J(S))- \frac{N k^{2}}{2 c} \tr\left(\partial_{x} S \partial_{x} S\right)+ N k \tr\left(\partial_{x} S E(S)\right)
 }
 \end{array}
 \eq
 and the Poisson brackets
 \beq\label{aa242}
 \begin{array}{c}
  \displaystyle{
\left\{S_{i j}(x), S_{k l}(y)\right\}=\frac{1}{N}\left(S_{i l}(x) \delta_{k j}-S_{k j}(x) \delta_{i l}\right) \delta(x-y)\,.
 }
 \end{array}
 \eq

 \paragraph{$U-V$ pair.}
 It is helpful to use the following notation
 \beq\label{aa243}
 \begin{array}{c}
  \displaystyle{
L(S,z)=\tr_{2} \Big(r_{12}(z) (1_N\otimes S) \Big)=NS_{0,0}E_1(z)1_N+
N \sum_{a\neq(0,0)} T_a S_a \varphi_a\left(z, \omega_a\right)\,,
 }
 \end{array}
 \eq
 where $\tr_2$ is the trace over the second tensor component and $r_{12}(z)$
 is the classical elliptic $r$-matrix (\ref{aa235}).
 The above expression, in fact, is the Lax matrix of the elliptic Euler-Arnold
 top \cite{LOZ,LOZ14} in finite-dimensional classical
 mechanics. Equations of motion for this model ${\dot S}=2c[S,J(S)]$ are obtained from (\ref{aa237}) in the
 limit $k\rightarrow 0$. It was also explained in  \cite{LOZ} that the $U$-matrix in the field theory case
 has the same form
 \beq\label{aa244}
 \begin{array}{c}
  \displaystyle{
U^{\hbox{\tiny{LL}}}(z)=L(S,z)
 }
 \end{array}
 \eq
 although here we imply $S=S(t,x)$, while in mechanics $S=S(t)$.
 For $V$-matrix we have
 \beq\label{aa245}
 \begin{array}{c}
  \displaystyle{
V^{\hbox{\tiny{LL}}}(z)= V_{1}(z)- c V_{2}(z)+\frac{N c^2 \vth'''(0)}{3\vth'(0)} 1_N\,,
 }
 \end{array}
 \eq
 where
 \beq\label{aa246}
 \begin{array}{c}
  \displaystyle{
V_1(z)= -N c \partial_z L(S,z)+ L(E(S)S,z)
 }
 \end{array}
 \eq
 and
 \beq\label{aa247}
 \begin{array}{c}
  \displaystyle{
V_{2}(z)=L\left(W, z\right)\,,\qquad W=-\frac{k}{c^{2}}\left[S, S_{x}\right]\,.
 }
 \end{array}
 \eq
The matrix $W$ is a solution of the equation $-k \partial_{x}S=\left[S, W \right]$.

\section{IRF-Vertex relation and change of variables}\label{sec4}
\setcounter{equation}{0}

\paragraph{Gauge equivalence.}
Let us introduce the following matrix:
 \beq\label{aa401}
 \begin{array}{c}
  \displaystyle{
G(z)=b(x, t) \Xi(z,q)D^{-1}(q)\in\Mat\,,
 }
 \end{array}
 \eq
 where the matrix $\Xi(z)$ and the diagonal matrix $D(q)$ are defined as follows:
 \beq\label{aa402}
 \begin{array}{c}
  \displaystyle{
\Xi_{i j}(z, q)=\theta \left[\begin{array}{c}
\frac{i}{N}-\frac{1}{2} \\
\frac{N}{2}
\end{array}\right]\left(z+N \bar{q}_{j} \mid N \tau\right)\,,
\quad \bar{q}_{j}=q_j-\frac{1}{N}\sum\limits_{k=1}^Nq_k\,,
 }
 \end{array}
 \eq
 \beq\label{aa403}
 \begin{array}{c}
  \displaystyle{
D_{i j}(q)=\delta_{i j} \prod_{k \neq i} \vartheta\left(q_{i}-q_{k}\right)\,.
 }
 \end{array}
 \eq
 The coefficient function $b(x,t)$ in (\ref{aa401}) has the form
 \beq\label{aa404}
 \begin{array}{c}
  \displaystyle{
b(x,t)=\prod\limits_{k<l}^N \vartheta\left(q_{l}-q_{k}\right)^{\frac{1}{N}}
\prod\limits_{m=1}^N\Big(N \left(k q_{m,x}-c\right) \Big)^{\frac{1}{2N}}\,.
 }
 \end{array}
 \eq
In the above formulae the condition (which is by definition of ${\bar q}_j$)
 \beq\label{aa405}
 \begin{array}{c}
  \displaystyle{
\sum_{k=1}^{N} \bar{q}_{k}=0
 }
 \end{array}
 \eq
is necessary. The defined above matrix $G(z)$
is (up to the function $b(x,t)$) the matrix of the IRF-Vertex transformation
\cite{Jimbo,LOZ,Z24}.

The reason why we use this matrix is as follows. The $U$-matrices for both models
have certain quasi-periodic behaviour in spectral parameter:
 \beq\label{aa406}
 \begin{array}{c}
  \displaystyle{
U^{\hbox{\tiny{2dCM}}}(z+1)=U^{\hbox{\tiny{2dCM}}}(z)\,,
}
\\ \ \\
  \displaystyle{
 U^{\hbox{\tiny{2dCM}}}(z+\tau)=\exp(2\pi\imath\,{\rm diag}(q_1,...,q_N))U^{\hbox{\tiny{2dCM}}}(z)
 \exp(-2\pi\imath\,{\rm diag}(q_1,...,q_N)-
}
\\ \ \\
  \displaystyle{
 -2\pi\imath c 1_N+2\pi\imath k\p_x{\rm diag}(q_1,...,q_N)
 }
 \end{array}
 \eq
 and
 \beq\label{aa407}
 \begin{array}{c}
  \displaystyle{
U^{\hbox{\tiny{LL}}}(z+1)=Q^{-1}U^{\hbox{\tiny{LL}}}(z)Q\,,
}
\\ \ \\
  \displaystyle{
 U^{\hbox{\tiny{LL}}}(z+\tau)=\Lambda^{-1} U^{\hbox{\tiny{LL}}}(z)}
 \Lambda-2\pi\imath c 1_N\,,
 \end{array}
 \eq
 where $Q,\Lambda$ are the matrices from (\ref{aa231}). The matrix $G(z)$ has very special
 structure. The action by $G(z)$ as in the gauge transformation
 \beq\label{aa408}
 \begin{array}{c}
   \displaystyle{
U^{\hbox{\tiny{LL}}}(z)=G(z) U^{\hbox{\tiny{2dCM}}}(z) G^{-1}(z)+k G_x(z) G^{-1}(z)
 }
 \end{array}
  \eq
  maps the quasi-periodic properties (\ref{aa406})
 into  (\ref{aa407}).
   On the one hand, the matrix $G(z)$ is degenerated at $z=0$:
 $\det G(0)=0$, that is $G^{-1}(z)$ has simple pole at $z=0$.
 On the other hand, the conjugation of $U^{\hbox{\tiny{2dCM}}}(z)$ by $G(z)$ does not provide
 the second order pole in $U^{\hbox{\tiny{LL}}}(z)$. Details can be found in \cite{LOZ}
 for a similar relation at the level of classical finite-dimensional mechanics.
See also \cite{VZ}, where different aspects of the transformation matrix $G(z)$ are discussed.

\paragraph{Change of variables.} The gauge transformation (\ref{aa408}) relates
both $U$-matrices. It allows to compute
explicit change of variables between the models. For any $a=(a_1,a_2)\in\mZ_N\times \mZ_N$ we have
 \beq\label{aa410}
 \begin{array}{c}
   \displaystyle{
 S_{a}(p,q,c)=\frac{c}{N}\, \delta_{a, (0,0)} +
 }
 \\ \ \\
    \displaystyle{
+(-1)^{a_{1}+a_{2}}\frac{e^{\pi\imath a_2\om_a}}{N^{2}} \left(\frac{\vartheta\left(\omega_{\alpha}\right)}{\vartheta^{\prime}(0)}\right)^N
\sum\limits_{m=1}^{N}
 \Big(P_{m}+c\sum\limits_{k:k\neq m}E_1(q_{mk}+\omega_a)\Big)
  \prod_{l: l \neq m}^{N} \varphi_a\left(q_{m}-q_{l}, \omega_a\right),
 }
 \end{array}
  \eq
  where $\om_a$ is from (\ref{aa234}) and
 \beq\label{aa411}
 \begin{array}{c}
   \displaystyle{
P_{m}=-p_{m}-\frac{k \alpha _{m, x}}{\alpha _m}+
 \sum\limits_{l: l\neq m}^{N} \alpha_{l}^{2} E_{1} \left( q_m-q_l\right)
 =
  }
 \\ \ \\
 \displaystyle{
 =
-p_{m}-\frac12\frac{k^2 q_{m,xx}}{kq_{m,x}-c}+
 \sum\limits_{l: l\neq m}^{N} (kq_{l,x}-c) E_{1} \left( q_m-q_l\right).
 }
 \end{array}
  \eq
  Using (\ref{aa2362}) one can write the formulae (\ref{aa410}) in the standard matrix basis as well.

\paragraph{Poisson map.} The statement that the obtained change of variables provides the Poisson map between
two models means that the canonical Poisson brackets (\ref{aa2031}) are mapped into the Lie-Poisson
brackets (\ref{aa242}). Put it differently, the Poisson brackets for $S_{ij}(p,q,c)$ computed
by means of (\ref{aa2031}) should reproduce (\ref{aa242}).

In fact, this statement was implicitly obtained in \cite{Z24}. It was proved that the field
analogue of the elliptic Calogero-Moser model is described by the non-ultralocal Maillet type
$r$-matrix structure
  \beq\label{aa412}
  \begin{array}{c}
  \displaystyle{
\{U^{\hbox{\tiny{2dCM}}}_1(z,x),U^{\hbox{\tiny{2dCM}}}_2(w,y)\}=
  }
  \\ \ \\
  \displaystyle{
\Big(-k\p_x {\bf r}_{12}(z,w|x)
+ [U^{\hbox{\tiny{2dCM}}}_1(z,x),{\bf r}^{\hbox{\tiny{2dCM}}}_{12}(z,w|x)]
-[U^{\hbox{\tiny{2dCM}}}_2(w,y),{\bf r}^{\hbox{\tiny{2dCM}}}_{21}(w,z|x)]\Big)\delta(x-y)-
  }
  \\ \ \\
  \displaystyle{
  -\Big({\bf r}^{\hbox{\tiny{2dCM}}}_{12}(z,w|x)+{\bf r}_{21}^{\hbox{\tiny{2dCM}}}(w,z|x)\Big)k\delta'(x-y)\,,
 }
 \end{array}
 \eq
where the classical $r$-matrix ${\bf r}^{\hbox{\tiny{2dCM}}}_{12}(z,w|x)$
 is very similar to its finite-dimensional version\footnote{Similar results are also known for the
 field analogue of the spin generalization of the Calogero-Moser systems and other 2d models \cite{Z}.}.
  At the same time the classical $r$-matrix
 structure for the Landau-Lifshitz model is
  \beq\label{aa413}
  \begin{array}{c}
  \displaystyle{
\{U^{\hbox{\tiny{LL}}}_1(z,x),U^{\hbox{\tiny{LL}}}_2(w,y)\}=
  }
  \\ \ \\
  \displaystyle{
\Big(
[U^{\hbox{\tiny{LL}}}_1(z,x),{\bf r}^{\hbox{\tiny{LL}}}_{12}(z,w|x)]
-[U^{\hbox{\tiny{LL}}}_2(w,y),{\bf r}^{\hbox{\tiny{LL}}}_{21}(w,z|x)]\Big)\delta(x-y)\,,
 }
 \end{array}
 \eq
where ${\bf r}^{\hbox{\tiny{LL}}}_{12}(z,w|x)$ is the elliptic non-dynamical $r$-matrix (\ref{aa235}).
It was shown in \cite{Z24} that the gauge transformation (\ref{aa408}) transforms (\ref{aa412})
into (\ref{aa413}). This is exactly what we need since (\ref{aa242}) follows from
(\ref{aa413}).

Existence of the classical $r$-matrix structures for both models means
the Poisson commutativity
$\{\tr(T^k(z,2\pi)),\tr(T^m(w,2\pi))\}=0$ for the corresponding monodromy matrices
 \beq\label{aa414}
 \begin{array}{c}
  \displaystyle{
 T(z,x)={\rm Pexp}\Big( \frac{1}{k}\int\limits_0^x dy\,  U(z,y) \Big)\,.
  }
 \end{array}
\eq
Due to the gauge equivalence the monodromies of both models are equal to each other.
This provides relation between the Hamiltonians.
Exact relation is as follows:
 \beq\label{aa415}
 \begin{array}{c}
  \displaystyle{
H^{\hbox{\tiny{2dCM}}}(x)=H^{\hbox{\tiny{LL}}}(x)-\frac{N^2 c^3 \vth'''(0)}{3\vth'(0)}\,.
  }
 \end{array}
\eq
 This relation was verified numerically. Its direct proof will be given elsewhere.

\subsection*{Acknowledgments}

This work was supported by the Russian Science Foundation under grant no. 25-11-00081,\\
https://rscf.ru/en/project/25-11-00081/ and performed at Steklov Mathematical Institute of Russian Academy of Sciences.

\section{Appendix: elliptic functions}\label{secA}
\def\theequation{A.\arabic{equation}}
\setcounter{equation}{0}

We actively use the elliptic Kronecker function:
 \beq\label{a01}
  \begin{array}{l}
  \displaystyle{
 \phi(z,u)=\frac{\vth'(0)\vth(z+u)}{\vth(z)\vth(u)}=\phi(u,z)\,,\quad
 \res\limits_{z=0}\phi(z,u)=1\,,
  \quad \phi(-z, -u) = -\phi(z, u)\,,
 }
 \end{array}
 \eq
where $\vth(z)$ is  theta-function:
 \beq\label{a02}
 \begin{array}{c}
  \displaystyle{
\vth(z)=\vth(z,\tau)\equiv-\theta{\left[\begin{array}{c}
1/2\\
1/2
\end{array}
\right]}(z|\, \tau )\,,
 }
 \end{array}
 \eq
\beq\label{a03}
 \begin{array}{c}
  \displaystyle{
\theta{\left[\begin{array}{c}
a\\
b
\end{array}
\right]}(z|\, \tau ) =\sum_{j\in \mZ}
\exp\left(2\pi\imath(j+a)^2\frac\tau2+2\pi\imath(j+a)(z+b)\right)\,,\quad {\rm Im}(\tau)>0\,.
}
 \end{array}
 \eq
By definition, $\vth(z)$ in (\ref{a02}) is
the first Jacobi theta function:
\beq\label{a031}
\begin{array}{c}
\displaystyle{
\theta_1(u|\tau )=\vth(u,\tau)=-i\sum_{k\in \mZ}
(-1)^k q^{(k+\frac{1}{2})^2}e^{\pi i (2k+1)u}\,,\quad q=e^{\pi i \tau}\,.
}
\end{array}
\eq
Relation to the standard Riemann and Jacobi notations is as follows:
  \beq\label{a121}
  \begin{array}{c}
    \displaystyle{
 \theta\left[\begin{array}{c}
1 / 2 \\
0
\end{array}\right](z, \tau)=
 \theta_{10}(z)=\theta_2(z)\,,\quad
 \theta\left[\begin{array}{c}
0 \\
0
\end{array}\right](z, \tau)=
 \theta_{00}(z)=\theta_3(z)\,,
   }
  \end{array}
   \eq
 $$
    \displaystyle{
 \theta\left[\begin{array}{c}
0\\
1/2
\end{array}\right](z, \tau)=
 \theta_{01}(z)=\theta_4(z)\,.
  }
  $$

The derivative $f(z,u) = \partial_u \phi(z,u)$ is given by
\beq\label{a04}
\begin{array}{c} \displaystyle{

    f(z, u) = \phi(z, u)(E_1(z + u) - E_1(u)), \qquad f(-z, -u) = f(z, u)
}\end{array}\eq
in terms of the first
  Eisenstein function:
\beq\label{a05}
\begin{array}{c} \displaystyle{
    E_1(z)=\frac{\vth'(z)}{\vth(z)}=\zeta(z)+\frac{z}{3}\frac{\vth'''(0)}{\vth'(0)}\,,
    \quad
    E_2(z) = - \partial_z E_1(z) = \wp(z) - \frac{\vartheta'''(0) }{3\vartheta'(0)}\,,
}\end{array}\eq
\beq\label{a06}
\begin{array}{c}
 \displaystyle{
    E_1(- z) = -E_1(z)\,, \quad E_2(-z) = E_2(z)\,,
}\end{array}
\eq
where $\wp(z)$ and $\zeta(z)$ are the Weierstrass functions. The second order derivative
$f'(z,u) = \partial^2_u \phi(z,u)$ is
\beq\label{a061}
\begin{array}{c}
 \displaystyle{
    f'(z,u)=\phi(z,u)\Big(\wp(z)-E_1^2(z)+2\wp(u)-2E_1(z)E_1(u)+2E_1(z+u)E_1(z)\Big)=
 }
 \\
 \displaystyle{
   =2\Big(\wp(u)-\rho(z)\Big)\phi(z,u)+2E_1(z)f(z,u) \,,
}\end{array}
\eq
where
\beq\label{a062}
\begin{array}{c}
\displaystyle{
\rho (z) = \frac{E^2_1(z) - \wp(z)}{2}\,.
}
\end{array}
\eq
 The defined above functions satisfy the widely known addition formulae:
\beq\label{a07}
  \begin{array}{c}
  \displaystyle{
  \phi(z_1, u_1) \phi(z_2, u_2) = \phi(z_1, u_1 + u_2) \phi(z_2 - z_1, u_2) + \phi(z_2, u_1 + u_2) \phi(z_1 - z_2, u_1)\,,
 }
 \end{array}
 \eq
\beq\label{a08}
  \begin{array}{c}
  \displaystyle{
 \phi(z,u_1)\phi(z,u_2)=\phi(z,u_1+u_2)\Big(E_1(z)+E_1(u_1)+E_1(u_2)-E_1(z+u_1+u_2)\Big)\,,
 }
 \end{array}
 \eq
\beq\label{a09}
  \begin{array}{c}
  \displaystyle{
  \phi(z, u_1) f(z,u_2)-\phi(z, u_2) f(z,u_1)=\phi(z,u_1+u_2)\Big(\wp(u_1)-\wp(u_2)\Big)\,,
 }
 \end{array}
 \eq
\beq\label{a10}
  \begin{array}{c}
  \displaystyle{
  \phi(z, u) \phi(z, -u) = \wp(z)-\wp(u)=E_2(z)-E_2(u)\,,
 }
 \end{array}
 \eq
\beq\label{a11}
  \begin{array}{c}
  \displaystyle{
  \phi(z, u) f(z, -u)-\phi(z, -u) f(z, u)=\wp'(u)\,.
 }
 \end{array}
 \eq
Also, the following two identities are useful:
\beq\label{a12}
  \begin{array}{c}
  \displaystyle{
   \frac{1}{2}\,\frac{\wp'(z)-\wp'(w)}{\wp(z)-\wp(w)}=\zeta(z+w)-\zeta(z)-\zeta(w)=
   E_1(z+w)-E_1(z)-E_1(w)
 }
 \end{array}
 \eq
and
\beq\label{a13}
  \begin{array}{c}
  \displaystyle{
   \Big(\zeta(z+w)-\zeta(z)-\zeta(w)\Big)^2=\wp(z)+\wp(w)+\wp(z+w)\,.
 }
 \end{array}
 \eq
%


\begin{small}

\end{small}


\begin{thebibliography}{99}
\addcontentsline{toc}{section}{References}


\bibitem{Krich22} A.A. Akhmetshin, I.M. Krichever, Y.S. Volvovski,
{\em Elliptic Families of Solutions of the Kadomtsev--Petviashvili Equation and the Field Elliptic Calogero--Moser System},
Funct. Anal. Appl., 36:4 (2002) 253–-266; arXiv:hep-th/0203192.

\bibitem{AtZ1} K. Atalikov, A. Zotov,
 {\em Field theory generalizations of two-body Calogero-Moser models in the form of Landau-Lifshitz equations},
J. Geom. Phys., 164 (2021) 104161; 	arXiv:2010.14297 [hep-th].

\bibitem{AtZ2}
K. Atalikov, A. Zotov,
 {\em Gauge equivalence between 1 + 1 rational Calogero–Moser field theory and higher rank Landau–Lifshitz equation},
JETP Letters, 117:8 (2023), 630--634; 	arXiv:2303.08020 [hep-th].

\bibitem{AtZ3}
K. Atalikov, A. Zotov,
 {\em Gauge equivalence of 1+1 Calogero-Moser-Sutherland field theory and higher rank trigonometric Landau-Lifshitz model},
 Theoret. and Math. Phys., 219:3 (2024), 1004–1017;	arXiv:2403.00428 [hep-th].

\bibitem{AtZ4} K. Atalikov, A. Zotov,
 {\em Higher rank 1+1 integrable Landau-Lifshitz field theories from associative Yang-Baxter equation},
JETP Lett. 115, 757-762 (2022); 	arXiv:2204.12576 [math-ph].


\bibitem{Baxter2}
R.J. Baxter,
 {\em Eight-vertex model in lattice statistics and
one-dimensional anisotropic Heisenberg chain. II. Equivalence to a
generalized ice-type lattice model},
Ann. Phys. 76 (1973) 25--47.

\bibitem{BD}
A. Belavin, V. Drinfeld,
{\em Solutions of the classical Yang–Baxter equation for simple Lie algebras},
Functional Analysis and Its Applications, 16:3 (1982) 159--180.

\bibitem{Calogero2}
%
F. Calogero,
{\em Exactly solvable one-dimensional many-body problems},
  Lett.
Nuovo Cim. 13 (1975) 411--416.


J. Moser,
{\em Three integrable Hamiltonian systems connected with isospectral deformations},
 Adv. Math. 16 (1975) 1--23.


M.A. Olshanetsky, A.M. Perelomov,
{\em  Classical integrable finite dimensional systems related to Lie algebras},
 Phys. Rep. 71 (1981) 313--400.

\bibitem{FT} L.D. Faddeev, L.A. Takhtajan,
{\em Hamiltonian methods in the theory of solitons},
Springer-Verlag, (1987).

\bibitem{Jimbo}
M. Jimbo, T. Miwa, M. Okado,
{\em Local state probabilities of solvable lattice models: An $A_{n-1}^{(1)}$ family},
Nuclear Physics B, 300 (1988) 74--108.

\bibitem{Kr}
I.M.  Krichever,
 {\em Elliptic solutions of the Kadomtsev--Petviashvili
equation and integrable systems of particles},
Funct.  Anal.  Appl., 14:4 (1980) 282--290.

\bibitem{Krich2} I. Krichever,
 {\em Vector bundles and Lax equations on algebraic curves},
Commun. Math. Phys., 229 (2002) 229–-269; arXiv:hep-th/0108110.

\bibitem{LL} L.D. Landau, E.M. Lifshitz,
 {\em To the Theory of Dispersion of Magnetic Permeability of Ferromagnetic Solids. Collection of L.D. Landau Works}, Nauka (1969), Vol. 1.
Phys. Zs. Sowjet., 8 (1935) 153--169.

\bibitem{LOZ} A. Levin, M. Olshanetsky, A. Zotov,
 {\em Hitchin Systems -- Symplectic Hecke Correspondence and Two-dimensional Version},
Commun. Math. Phys.  236 (2003) 93--133;     arXiv:nlin/0110045.

A.V. Zotov, A.V. Smirnov,
{\em Modifications of bundles, elliptic integrable systems, and related problems},
 Theoret. and Math. Phys., 177, (2013) 1281--1338.


\bibitem{LOZ14} A. Levin, M. Olshanetsky, A. Zotov,
{\em Relativistic Classical Integrable Tops and Quantum R-matrices},
JHEP 07 (2014) 012,
arXiv:1405.7523 [hep-th].

A. Levin, M. Olshanetsky, A. Zotov,
 {\em Noncommutative extensions of elliptic integrable Euler-Arnold tops and Painleve VI equation},
J. Phys. A: Math. Theor. 49:39 (2016) 395202; arXiv:1603.06101 [math-ph].








\bibitem{STS} A.G. Reiman, M.A. Semenov-Tian-Shansky,
 {\em Lie algebras and Lax equations with spectral parameter on an elliptic curve},
Zap. Nauchn. Sem. LOMI, 150 (1986) 104–-118.


A.V. Zotov,
 {\em 1+1 Gaudin Model},
SIGMA 7 (2011), 067; 	arXiv:1012.1072 [math-ph].







\bibitem{Skl} E.K. Sklyanin,
{\em On complete integrability of the Landau-Lifshitz equation},
Preprint LOMI, E-3-79, Leningrad (1979).

E. K. Sklyanin, {\em On the Poisson structure of the periodic classical XYZ-chain},
 Questions of quantum field theory and statistical physics. Part 6,
 Zap. Nauchn. Sem. LOMI, 150, (1986) 154--180.


\bibitem{VZ} M. Vasilyev, A. Zotov, {\em On factorized Lax pairs for classical many-body integrable systems},
	Reviews in Mathematical Physics, 31:6 (2019) 1930002; 	arXiv:1804.02777 [math-ph]


\bibitem{ZZ} A. Zabrodin, A. Zotov,
{\em Field analogue of the Ruijsenaars-Schneider model},
JHEP 07 (2022) 023;	arXiv: 2107.01697 [math-ph].



\bibitem{ZaSh} V.E. Zakharov, A.B. Shabat,
 {\em Exact Theory of Two-dimensional Self-focusing and One-dimensional Self-modulation of Waves in Nonlinear Media},
 Soviet physics JETP 34:1 (1972) 62--69.

V.E. Zakharov, A.B. Shabat,
 {\em A scheme for integrating the nonlinear equations
of mathematical physics by the method of the inverse scattering problem. I}
Funct. Anal. Appl., 8:3 (1974), 226–-235.

V.E. Zakharov, A.B. Shabat,
 {\em Integration of nonlinear equations of mathematical physics by the method of inverse scattering. II}
Funct. Anal. Appl., 13:3 (1979), 166–-174.






\bibitem{Z24} Andrei Zotov,
{\em Non-ultralocal classical r-matrix structure for 1+1 field analogue of elliptic Calogero–Moser model},
  J. Phys. A, 57 (2024), 315201;
	arXiv:2404.01898 [hep-th].



\bibitem{Z}
A. Zotov,
{\em On the field analogue of elliptic spin Calogero-Moser model: Lax pair and equations of motion},
Funktsional. Anal. i Prilozhen., 59:2 (2025), 46--66;
arXiv:2407.13854 [nlin.SI].

D. Domanevsky, A. Levin, M. Olshanetsky, A. Zotov,
{\em Integrable deformations of principal chiral model from solutions of associative Yang-Baxter equation},
Izvestiya: Mathematics (2026) to appear; arXiv:2501.08777 [math-ph].

A. Levin, M. Olshanetsky, A. Zotov,
{\em 2d Integrable systems, 4d Chern-Simons theory and Affine Higgs bundles},
Eur. Phys. J. C 82, 635 (2022); arXiv:2202.10106 [hep-th].

\end{thebibliography}
\end{document}